# Adaptive Neighborhood Graph Construction for Inference in Multi-Relational Networks

Shobeir Fakhraei[1,2]  Dhanya Sridhar[2]  Jay Pujara[2]  Lise Getoor[2]
shobeir@cs.umd.edu  dsridhar@ucsc.edu  jay@cs.umd.edu  getoor@soe.ucsc.edu

[1]University of Maryland, College Park, MD, USA
[2]University of California, Santa Cruz, CA, USA

## ABSTRACT

A *neighborhood graph*, which represents the instances as vertices and their relations as weighted edges, is the basis of many semi-supervised and relational models for node labeling and link prediction. Most methods employ a sequential process to construct the neighborhood graph. This process often consists of generating a *candidate graph*, pruning the candidate graph to make a neighborhood graph, and then performing inference on the variables (i.e., nodes) in the neighborhood graph. In this paper, we propose a framework that can dynamically adapt the neighborhood graph based on the states of variables from intermediate inference results, as well as structural properties of the relations connecting them. A key strength of our framework is its ability to handle multi-relational data and employ varying amounts of relations for each instance based on the intermediate inference results. We formulate the link prediction task as inference on neighborhood graphs, and include preliminary results illustrating the effects of different strategies in our proposed framework.

## 1. INTRODUCTION

Neighborhood graphs which capture interdependencies between instances, are the underlying structure of reasoning in many *graph-based* predictive models. These models are used in domains such as collaborative filtering, link prediction, and image classification. For example, when determining characteristics of individuals, the characteristics of the people who are most similar to them, their friends, family, and co-workers, can all influence a model's prediction. In these models the data is represented as a *neighborhood graph* or network, where nodes are instances and weighted edges represent relations (e.g., similarities) between them. Neighborhood graph-based methods include popular semi-supervised modeling techniques.

Most methods that make predictions based on a neighborhood graph can be characterized in terms of three basic operations [1]: *Candidate graph Generation*, *Selection* and *Inference*. The first step, is the candidate graph *generation*, which often includes defining the relations or similarities between instances. The process of constructing the candidate graph is generally problem-specific. If the original input data is *relational*, or in the form of a graph (which we call *data graph*), some of the explicit relations such as relationships in a social network, or adjacencies in an image may be used as an approximation of the affinity or dependency of instances. When the original input data includes instance attributes, a similarity or kernel function is defined to estimate the pairwise affinity of the items. The abundance of pairwise similarities or relations often hinders a model's scalability as well as its predictive performance and makes the candidate graph generally unsuitable for modeling approaches.

The next step is *selection*, reducing the candidate graph by pruning similarities or relations to a more manageable neighborhood graph. Examples of these methods that are often considered a pre-processing step to inference include k-nearest neighbors, $\epsilon$-neighborhood selection, and b-matching [1]. This is an important step in all the neighborhood graph-based methods as unnecessary relations in the graph reduces the scalability. More importantly, similar to the negative effect of a large $k$ in a simple $k$-nearest neighbors classifier, unnecessary relations can harm the performance of a neighborhood graph-based model. Fakhraei et al. [2] show this negative effect in a drug target prediction setting.

The *third step* is the algorithm to perform inference using the neighborhood graph. Methods such as Mincut, graph random walk, Gaussian random fields, local and global consistency, spectral graph transducer, manifold regularization, and label propagation are examples that perform inference on the neighborhood graph [3].

We consider several challenges in this sequential process for constructing the neighborhood graph especially based on *multi-relational* data:

– Most methods are designed to handle a single similarity, dependency, or relation type when constructing the neighborhood graph. However, there are often multiple relations than each can serve as a noisy approximation for the affinity that is important for the predictive task at hand. Constructing a *multi-relational* neighborhood graph that can effectively combine different affinity signals from multiple sources is highly important.

– The model-agnostic nature of the pruning methods can result in neighborhood graphs that may not adequately capture the most important similarities for a particular model. We assert that successful graph construction requires a model-based approach that includes the context of the prediction problem as one of its components. In other words, model-aware graph construction methods can leverage the information about the uncertainties and decision boundaries of the model that sequential model-agnostics methods can not.

– The pruning process often solely relies on the value assigned to the similarity or relation and does not consider the characteristics of instances that are being connected in the neighborhood graph. For example, a relatively low similarity to an instance of a rare class may be more important than a high similarity to an instance of the majority class.
– With the rapid growth in size of the datasets, performing complex search processes on all pairwise relations between instances increases the computational cost. Therefore, depending on the size of the dataset an algorithm may only be able to make approximate decisions based on partial observations.

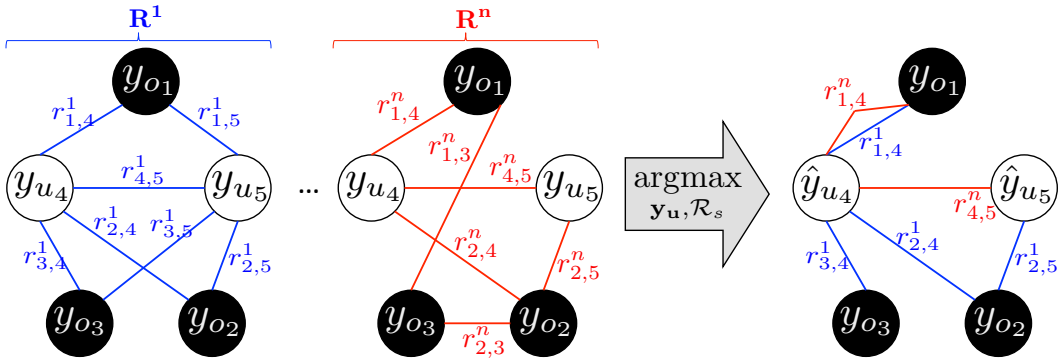

Figure 1: Neighborhood graph construction from candidate graph with several relations. For simplicity of notation, items $\mathbf{x}_i$ are represented only with their labels $y_i$. Observed labels $y_o$ are shown as black circles and instances with unobserved labels $y_u$ are shown with white circles and annotated accordingly. Relations $r_{\mathbf{x}_i,\mathbf{x}_j}$ are noted as $r_{i,j}$. The neighborhood graph shown on the right has a subset of selected relations with different types and also the inferred values ($\hat{y}$) for the unobserved labels.

In this paper, we address these challenges by developing a general framework for dynamically constructing a neighborhood graph. Our framework enables rich models for *active inference* that interleave inference and neighborhood graph construction, efficiently using multi-relational data while maintaining scalable performance. We have developed the **LINA** framework, which consists of four parts; **L**earning the relative importance of each relation in multi-relational settings, **I**nferring labels for interrelated instances, **N**ominating instances that can benefit from additional relations in the neighborhood graph, and **A**ctivating new relations between instances in the neighborhood graph to improve the inference results. Our main contribution in this paper include:

- We propose a general unified framework to actively learn multi-relational neighborhood graphs and demonstrate its use for collective link prediction.
- We explore the ramifications of modeling choices for the important components in our framework and their combination; Nomination and Activation.
- We lay out a general formulation for the link prediction task as an inference problem on a neighborhood graph.
- We present preliminary results of applying different components of our framework for link prediction on a drug-target interaction network dataset.

In Section 2 we formally define the problem statement, and in Section 3 we describe our proposed framework. We explain *Hinge-loss Markov random fields* that we use for inference in Section 4. We then present our nomination, activation, and learning methods in Section 5. Section 6 includes discussion on modeling the link prediction task with a neighborhood graph. We then provide preliminary experimental results of applying our approach in Section 7.

## 2. GENERAL PROBLEM STATEMENT

Let candidate graph be represented via $\mathcal{G}_c \triangleq \langle \mathbf{X}, \mathcal{R} \rangle$, where vertices $\mathbf{X}$ is a set of $n$ data points $\mathbf{x}_i = [x_i^1, \ldots, x_i^{p-1}, y_i]$, and $\mathbf{y}$ represent the set of all labels ($\mathbf{y} = \{y_1, \ldots, y_n\}$) where a subset is observed ($\mathbf{y_o}$) and others are unobserved ($\mathbf{y_u}$), and edges $\mathcal{R}$ are a set of relations $r_{\mathbf{x}_i,\mathbf{x}_j}$ (e.g., pairwise similarities, or connections in social network) connecting pairs of data points $\mathbf{x}_i$ and $\mathbf{x}_j$, based on some notion of affinity. Graph-based method generally make the assumption that for two data points $\mathbf{x}_i$ and $\mathbf{x}_j$ that are connected in this candidate graph $\mathcal{G}_c$, their labels $y_i$ and $y_j$ are close to each other, where the strength of this assumption depends on the value or weight associated with the relation $r_{\mathbf{x}_i,\mathbf{x}_j}$.

Models that perform collective inference [4] based on both known and unknown labels find an optimal state of all unknowns variables by optimizing an objective function $f$ over all target variables $\mathbf{y_u}$ and jointly assigning values to all of them. Thus, collective methods propagate inferred values of the labels based on the relations on the candidate graph. Formally:

$$\hat{\mathbf{y}}_\mathbf{u} = \arg \operatorname*{opt}_{\mathbf{y_u}} f(\mathbf{y_u}, \mathbf{y_o}, \mathcal{G}_c; \boldsymbol{\omega})$$

e.g., in a probabilistic setting:

$$\hat{\mathbf{y}}_\mathbf{u} = \arg \max_{\mathbf{y_u}} \mathrm{P}(\mathbf{y_u}|\mathbf{y_o}, \mathcal{G}_c; \boldsymbol{\omega})$$

where, $\boldsymbol{\omega} = \{\omega_1, \ldots, \omega_m\}$ represents the model parameters.

Furthermore, we are interested in settings where $\mathcal{R}$ can be of different types (i.e., $\mathcal{R} = \{\mathbf{R^1}, \ldots, \mathbf{R^n}\}$), for example we can have pairwise similarities computed based on different methods or features, or observed relations in a social network with different semantic (friendship, follow, etc.).

The task of interest in this paper is given a fixed *activation quota* $q \geq 1$ to dynamically select a subset $\mathcal{R}_s$ (such that $|\mathcal{R}_s| \leq q$) from all known relations $\mathcal{R}$ to improve the performance and scalability of inference. We call the reduced candidate graph with less relations a *neighborhood graph* ($\mathcal{G}_n$). We select some of the relations $r_{\mathbf{x}_i,\mathbf{x}_j}$ from each relation type $\mathbf{R}^k$ form candidate graph ($\mathcal{G}_c$) to include in the neighborhood graph ($\mathcal{G}_n$) and discard the rest of relations. Figure 1 shows an overview of our main task.

More formally we aim to find an *activation method* $\tau$ (i.e., inclusion function or map) such that $\mathcal{R}_s = \tau(\mathcal{R}, \mathbf{y}, q_\tau, \boldsymbol{\omega})$ and $|\mathcal{R}_s| \leq q_\tau$, where using a graph $\mathcal{G}_n \triangleq \langle \mathbf{X}, \mathcal{R}_s \rangle$ with $\mathcal{R}_s$ instead of $\mathcal{R}$ improves the inference's result, i.e.,

$$\hat{\mathbf{y}}_\mathbf{u} = \underset{\mathbf{y}_\mathbf{u}, \mathcal{R}_s}{\arg\max}\, P(\mathbf{y}_\mathbf{u}|\mathbf{y}_\mathbf{o}, \langle \mathbf{X}, \mathcal{R}_s \rangle; \boldsymbol{\omega}) \qquad (1)$$

Next, we describe our proposed methods to maximize the objective in (1).

## 3. PROPOSED GENERAL FRAMEWORK

Maximizing the objective in (1) is a non-convex combinatorial optimization problem. Hence, we break this objective into three parts; one part to *nominate* instance ($\mathbf{X}_\mathbf{n}$) with unknown labels that require more evidence, another part to *activate* or select more relations for those nominated instances ($\mathcal{R}_s$), and the third part to jointly *infer* all unknown labels for instances given the results from the two previous steps.

To achieve this we introduce a *nomination method* $\eta$ such that $\mathbf{X}_\mathbf{n} = \eta(\mathbf{X}, \hat{\mathbf{y}}_\mathbf{u}, q_\eta)$ and $|\mathbf{X}_\mathbf{n}| \leq q_\eta$ where $q_\eta$ indicates the *nomination quota*. Note that nomination and activation method each have a quota of their own that we can tune. For simplicity we assume that the *activation quota* is related to the *nomination quota* by a constant $\kappa$ (i.e. $q_\tau = \kappa \times q_\eta$), which means for each nominated instance we can activate up to a maximum of $\kappa$ relations $r$. We then modify the activation method to depend on the nominated instances such that $\mathcal{R}_s = \tau(\langle \mathbf{X}_n, \mathcal{R} \rangle, \mathbf{y}, q_\tau, \boldsymbol{\omega})$

Then (1) will be approximated via three components of nomination, activation, and inference as following:

$$\mathbf{X}_\mathbf{n} = \eta(\mathbf{X}, \hat{\mathbf{y}}_\mathbf{u}, q_\eta)$$
$$\mathcal{R}_s = \tau(\langle \mathbf{X}_n, \mathcal{R} \rangle, \mathbf{y}, q_\tau, \boldsymbol{\omega})$$
$$\hat{\mathbf{y}}_\mathbf{u} = \underset{\mathbf{y}_\mathbf{u}}{\arg\max}\, P(\mathbf{y}_\mathbf{u}|\mathbf{y}_\mathbf{o}, \langle \mathbf{X}, \mathcal{R}_s \rangle; \boldsymbol{\omega})$$

Due to obvious dependencies between $\mathbf{X}_\mathbf{n}, \mathcal{R}_s$, and $\hat{\mathbf{y}}_\mathbf{u}$, we developed an iterative algorithm to preform these steps and update the assignments. Algorithm 1 shows the overall iterative code performing each step.

---
**Algorithm 1** LINA Framework
---
1: $\mathcal{R}_s \leftarrow \mathcal{R}_{\text{init}}$
2: $\boldsymbol{\omega} \leftarrow$ Learn the model parameters based on $\mathcal{R}_{\text{init}}$
3: $\mathcal{R}_s \leftarrow \emptyset$
4: **for** $i \in 0 \ldots$ total iterations **do**
5: $\quad \mathbf{y}_\mathbf{u} \leftarrow \arg\max P(\mathbf{y}_\mathbf{u}|\mathbf{y}_\mathbf{o}, \langle \mathbf{X}, \mathcal{R}_s \rangle; \boldsymbol{\omega})$
6: $\quad \mathbf{X}_\mathbf{n} \leftarrow \eta(\mathbf{X}, \mathbf{y}_\mathbf{u}, q_\eta)$
7: $\quad \mathcal{R}_s \leftarrow \tau(\langle \mathbf{X}_n, \mathcal{R} \rangle, \mathbf{y}, q_\tau, \boldsymbol{\omega})$
8: **return** $\mathbf{y}_\mathbf{u}$
---

We introduce a set of approaches for *nomination method* $\tau$ and *activation method* $\eta$ described in Section 5. We also discuss an approach to learn the model parameters $\boldsymbol{\omega}$, as weights that capture the importance of each relation type $\mathbf{R}^k$ in Section 4.2.

## 4. HINGE-LOSS MRFS

The methods introduced in this paper generally apply to most neighborhood graph-based probabilistic models that perform collective inference on all unknown variables. One particular model of interest in this paper is an instance of continuous-valued Markov random field models (MRFs) with a strongly convex MAP inference objective function, known as *hinge-loss Markov random fields* (HL-MRFs) [5]. An HL-MRF is a continuous-valued Markov network in which the potentials are hinge functions of the variables. Our choice of HL-MRFs comes from technical considerations: MAP inference in HL-MRFs is provably and empirically efficient, in theory growing $O(N^3)$ with the number of potentials, $N$, but in practice often converging in $O(N)$ time. Models built using HL-MRFs have achieves state-of-the-art performance for a variety of applications including drug target prediction [2], drug interaction prediction [6] recommender systems [7], student engagement analysis [8], knowledge graph identification [9], and social spammer detection [10]. Finally, HL-MRFs are easily specified through *probabilistic soft logic* (PSL) [5], a probabilistic programming language with a first-order logic-like syntax.

A hinge-loss MRF defines a joint probability density function of the form

$$P(\mathbf{y}_\mathbf{u}|\mathbf{X}, \mathbf{y}_\mathbf{o}) = \frac{1}{\mathcal{Z}} \exp\left(-\sum_{r=1}^{M} \omega_r \phi_r(\mathbf{y}_\mathbf{u}, \mathbf{X}, \mathbf{y}_\mathbf{o})\right), \qquad (2)$$

where the entries of target variables $\mathbf{y}_\mathbf{u}$ and observed variables $\mathbf{X}$ and $\mathbf{y}_\mathbf{o}$ are in $[0,1]$, $\boldsymbol{\omega}$ is a vector of weight parameters, $\mathcal{Z}$ is a normalization constant, and

$$\phi_r(\mathbf{y}_\mathbf{u}, \mathbf{X}, \mathbf{y}_\mathbf{o}) = (\max\{l_r(\mathbf{y}_\mathbf{u}, \mathbf{X}, \mathbf{y}_\mathbf{o}), 0\})^{\rho_r} \qquad (3)$$

is a *hinge-loss potential* specified by a linear function $l_r$ and optional exponent $\rho_r \in \{1, 2\}$. Relaxations of first-order logic rules are one way to derive the linear functions $l_r(\cdot)$ in the hinge-loss potentials $\phi_r$. Thus, a set of logical rules described in the PSL framework is a template for an HL-MRF model. Given a collection of logical implications based on domain knowledge described in PSL and a set of observations from data, the rules are instantiated, or grounded out, with known entities in the dataset. Each instantiation of the rules maps to a hinge-loss potential function as in (3), and the potential functions define the HL-MRF model.

To illustrate modeling in PSL, we consider a similarity-based rule that encourages transitive closure for link prediction between entities $a, b,$ and $c$:

$$\text{SIMILAR}(a,b) \land \text{LINK}(b,c) \rightarrow \text{LINK}(a,c)$$

where instantiations of the predicate LINK represent continuous target variables for a link prediction task and instantiations of SIMILAR are continuous observed variables. The convex relaxation of this logical implication derived using the well-known Lukasiewicz logic for continuous truth values is equivalent to the hinge-loss function

$$\max(\text{SIMILAR}(a,b) + \text{LINK}(b,c) - \text{LINK}(a,c) - 1, 0)$$

and can be understood as its *distance to satisfaction*. The distance to satisfaction of this ground rule is a linear function of the variables and thus, exactly corresponds to

$$\phi_r(\text{LINK}(b,c), \text{LINK}(a,c), \text{SIMILAR}(a,b))$$

the feature function that scores configurations of assignments to the three variables. Intuitively, distance to satisfaction represents the degree to which the rule is violated by assignments to the random variables conditioned on the observations. We describe MAP inference and parameter estimation in HL-MRF models below. Intuitively, MAP inference minimizes the weighted, convex distances to satisfaction to find an consistent joint assignment for all the target variables and the weight parameters convey relative importance of each rule by varying the penalty for violating that rule.

## 4.1 MAP Inference

We perform MAP inference in HL-MRFs to find the best assignment to all target variables given evidence. Formally, the MAP inference objective is of the form

$$\arg\max_{\mathbf{y_u} \in [0,1]^n} \frac{1}{\mathcal{Z}} \exp\left(-\sum_{r=1}^{M} \omega_r \phi_r(\mathbf{y_u}, \mathbf{X}, \mathbf{y_o})\right) \quad (4)$$

$$\equiv \arg\min_{\mathbf{y_u}} \sum_{r=1}^{m} \omega_r \max\{l_r(\mathbf{y_u}, \mathbf{X}, \mathbf{y_o}), 0\} \quad (5)$$

HL-MRFs has an advantage over other Markov networks since the MAP problem can be solved exactly and in polynomial time as a convex optimization problem. There are many off-the-shelf convex optimization solvers such as interior-point methods, but here we use the notable Alternating Direction Method of Multiples algorithm (ADMM) [11]. The ADMM algorithm uses consensus optimization to divide the MAP problem into independent subproblems. For full details on consensus optimization with ADMM for HL-MRF MAP inference, see [5].

## 4.2 Maximum Likelihood Parameter Learning

Each logical rule in PSL that templates a set of hinge-loss potentials when ground out has an associated weight $\omega_r$. The vector of weights $\boldsymbol{\omega}$ are the parameters of an HL-MRF model and can be learned from training data. The canonical approach for parameter estimation is maximum likelihood estimation (MLE) which maximizes the log-likelihood of the training data. The partial derivative of an HL-MRF with respect to any $\omega_r$ is

$$\frac{\partial \log P(\mathbf{y_u}|\mathbf{X}, \mathbf{y_o})}{\partial \omega_r} = \\ \mathbb{E}_{\boldsymbol{\omega}}\left[\sum_{j \in g_r} \phi_r(\mathbf{y_u}, \mathbf{X}, \mathbf{y_o})\right] - \sum_{j \in g_r} \phi_r(\mathbf{y_u}, \mathbf{X}, \mathbf{y_o}) \quad (6)$$

where $g_r$ are the groundings of rule $r$ on the training data and $\mathbb{E}_{\boldsymbol{\omega}}$ is the expectation of the HL-MRF distribution parameterized by $\boldsymbol{\omega}$. Intuitively, the gradient with respect $\omega_r$ compares the expected sum of the potentials defined by $r$ to the actual sum based on training data. Smaller gradients indicates better fit to the training data. Gradient descent is performed using the structured voted perceptron algorithm [5]. However, as in other joint models, $\mathbb{E}_{\boldsymbol{\omega}}$ is intractable to compute and so we use a common approximation, the values of the potential functions $\phi_r$ at the MAP state.

## 5. METHODS

We use HL-MRFs for inference and parameter learning of our framework. In this section we discuss our approach for nomination and activation method and explain how we leverage the optimization terms and model parameters in our proposed methods.

## 5.1 Nominating Instances

The nomination phase of our framework selects instances that may benefit from additional evidence. The process of nominating instances is similar to the problem of active learning [12], where instances are labeled based on a utility function. However, in contrast to active learning, nomination does not acquire labels, but selects those instances for which we introduce new relations in the neighborhood graph.

Our general framework is compatible with arbitrary nomination techniques, allowing us to leverage the diverse active learning strategies developed over the past decades. In addition, our choice of HL-MRFs for modeling inference problems provides the opportunity to use unique nomination strategies that incorporate partial inference outputs and the model state. Here, we present nomination strategies that use inference context and model features to choose instances.

### 5.1.1 Model-Aware Nomination

Probabilistic models that use a neighborhood graph define a rich set of relations between instances that can provide useful structural features when nominating instances. We introduce nomination methods that use these structural features to provide a model-aware nomination method.

We build on a method from Pujara et al. [13], which derives features for instance selection from the optimization process underlying inference. In prior work, these features were used in an online inference setting, where instances were selectively updated in response to new evidence. In our setting, we use these model features to determine which instances would benefit from additional evidence.

Pujara et al. [13] observe that model structure (in our setting, relations between instances in the neighborhood graph) translate directly into optimization terms in the inference objective. Features from these optimization terms allow model-based scoring of instances that are difficult to optimize, and thus might benefit from additional evidence. Moreover, since the optimization is central to inference, these features can be generated with little or no overhead.

The methods we present identify features from the popular consensus optimization algorithm, the alternating direction method of multipliers (ADMM) [11]. ADMM decomposes the optimization objective into independent subproblems, optimizes each subproblem independently, and introduces a constraint that the all subproblems agree on the optimal value of each inferred variable. This optimization is often expressed using the augmented Lagrangian seen in (7). Here, $\omega_r$ and $\phi_r$ are the parameter and potential associated with a given relation $r$, $\tilde{y}_r$ is the local optimizer of a subproblem and $y_r$ is the consensus estimate. Consensus between subproblems is enforced by introducing a Lagrange multiplier, $\alpha_r$, associated with the constraint, and increasing the optimization penalty, $\rho$, associated with violating this constraint to guarantee eventual convergence.

$$\min_{\tilde{y}_r} \omega_r \phi_r(\mathbf{x}, \tilde{y}_r) + \frac{\rho}{2} \left\| \tilde{y}_r - y_r + \frac{1}{\rho} \alpha_r \right\|^2 \quad (7)$$

A useful intuition is that instances where existing relations cause disagreement on a label are useful candidates for nomination. This intuition can be expressed in terms of the Lagrange multipliers associated with each optimization term. At convergence, the value of this Lagrange multiplier captures the disagreement of a given optimization term with the consensus estimate. Thus, by nominating instances as-sociated with a potential with high Lagrange multipliers, we can improve our estimate of controversial instances. We use an uncertainty measures based on the Lagrange multipliers from ADMM optimization. The average weighted Lagrange multiplier (AWL) [13], measures the overall discrepancy between the local and consensus copies of the label:

$$\frac{1}{|R|}\sum_{r\in R}\omega_r \alpha_r(i) \quad (8)$$

where $R$ indicates all the local copies of a consensus variable.

## 5.2 Activating Relations

In the relation activation step, we select a subset of relations $\mathcal{R}_s$ from the candidate graph $\mathcal{G}_c$ to include in the neighborhood graph $\mathcal{G}_n$. Formally, the activated relations $\mathcal{R}_s = \tau(\mathbf{X}_n, \mathcal{G}_n, q_\tau, \boldsymbol{\omega})$ are chosen by function $\tau$ given the nominated instances and the set of all relations $\mathcal{R}$. In addition to considering the weight, or value, of each relation edge $r^k_{\mathbf{x}_i,\mathbf{x}_j}$, we design $\tau$ to rank relations based on structural properties of the neighborhood graph. At a high level, we use multiple features to score each $r^k_{\mathbf{x}_i,\mathbf{x}_j}$ and select the top $q_\tau$ relations based on their combined scores. In addition to the edge weight feature, we introduce features that use $x_\alpha \in \mathcal{G}_n$ that are incident to $r^k_{\mathbf{x}_i,\mathbf{x}_j}$. Intuitively, these structural features measure the informativeness of a relation for inferring multiple unknown instances, and its ability to effectively propagate evidence through $\mathcal{G}_n$.

First we introduce and define two additional structural features along with the relation value feature. Then, we fully describe how $\tau$ combines these features and selects the top $q_\tau$ relations from multiple relation types.

### 5.2.1 Value Feature

We use strength or value associated with a relation edge $r^k_{\mathbf{x}_i,\mathbf{x}_j}$ as a basic feature. Relations of higher value convey a greater dependence between assignments to labels of instances $\mathbf{x_i}$ and $\mathbf{x_j}$. If $\mathbf{x_i}$ or $\mathbf{x_j}$ is an instance with known label, then a high valued $r^k_{\mathbf{x}_i,\mathbf{x}_j}$ effectively propagates that label to the unknown instance.

### 5.2.2 Nominated Instance Count Feature

For each $r^k_{\mathbf{x}_i,\mathbf{x}_j}$ in $\mathcal{G}_n$, we compute the number of nominated unknown instances $\mathbf{X}_n$ that are incident upon $r^k_{\mathbf{x}_i,\mathbf{x}_j}$. We require the use of an incident operator $\mathcal{I}(x_\alpha, r_\beta)$ that returns 1 if $x_\alpha$ shares an endpoint with $r^k_{\mathbf{x}_i,\mathbf{x}_j}$ and 0 otherwise.

Formally, the nominated instance count score for $r^k_{\mathbf{x}_i,\mathbf{x}_j}$ is:

$$\sum_{\mathbf{x}\in\mathbf{X}_n} \mathcal{I}\left(\mathbf{x}, r^k_{\mathbf{x}_i,\mathbf{x}_j}\right)$$

where we only consider nominated unknown instances incident to $r^k_{\mathbf{x}_i,\mathbf{x}_j}$. Intuitively, if many $\mathbf{x} \in \mathbf{X}_n$ have endpoints in $r^k_{\mathbf{x}_i,\mathbf{x}_j}$, then the relation will be informative for many predictions and introduce multiple useful dependencies in the inference step.

### 5.2.3 Observed Instance Count Feature

For each relation $r^k_{\mathbf{x}_i,\mathbf{x}_j}$, we also compute the number of instances with *observed* labels, instances with $\mathbf{y}_o$ that are incident to $r^k_{\mathbf{x}_i,\mathbf{x}_j}$. Observed links are important because they propagate evidence to unknown link instances through the activated relations. Formally, this feature score for $r^k_{\mathbf{x}_i,\mathbf{x}_j}$ is:

$$\sum_{\mathbf{x},y|y\in\mathbf{y_o}} \mathcal{I}\left(\mathbf{x}, r^k_{\mathbf{x}_i,\mathbf{x}_j}\right)$$

where we only consider observed instances incident to $r^k_{\mathbf{x}_i,\mathbf{x}_j}$. Higher valued $r^k_{\mathbf{x}_i,\mathbf{x}_j}$ potentially connect nominated unknown instances $\mathbf{X}_n$ to many instances with $\mathbf{y_o}$. The observed instances provide valuable evidence for predictions of the unknowns.

### 5.2.4 Combining Features and Selecting Relations with $\tau$

Finally, we require selection function $\tau$ that uses the proposed features over relations to select the most useful $q_\tau$ as evidence for predictions of $\mathbf{X}_n$. In our work, for each $\mathbf{x}_i \in \mathbf{X}_n$ we consider the set of relations $\mathcal{R}_i = \{r_\beta | \mathcal{I}(\mathbf{x}_i, r_\beta)\}$ to which unknown instance $\mathbf{x}_i$ is incident on. For each $r^k_{\mathbf{x}_j,\mathbf{x}_l} \in \mathcal{R}_i$, we compute scores for each feature and take the product of scores, which we denote $s^k_{\mathbf{x}_j,\mathbf{x}_l}$. We rank $\mathcal{R}_i$ by $s^k_{\mathbf{x}_j,\mathbf{x}_l} \times \omega_k$ where $\omega_k$ is the *parameter*, or importance, of relation type $k$ learned in Section 4.2. For each $\mathbf{x}_i$, we select the top $\kappa$ relations from $\mathcal{R}_i$. Since $q_\tau = \kappa \times q_\eta$, where $q_\eta$ is the number of nominated instances, the activation quota is never exceeded.

## 6. GRAPH-BASED LINK PREDICTION

While the framework proposed in the paper is generally applicable to all neighborhood graph-based models, we focus our arguments on the link prediction task. Inferring information about links is the basis for many machine learning and data science tasks. Link prediction, such as predicting which people would become friends on a social networks or which authors will cite each other in a scholar network; recommender systems, such as predicting which article or item is more relevant to a user; or biological predictions, such as which two drugs will interact with each other, or which drug will interact with a protein are examples of link predictionin different networks. When the inferred labels are binary such as click prediction the task is often called *link prediction*, and when the label is continuous or multi-valued such as ratings prediction the closely related task is often called *link regression* [14].

To apply a neighborhood graphbased learning method for a link prediction task, the nodes $\mathbf{X}$ in the candidate graph $\mathcal{G}_c \triangleq \langle \mathbf{X}, \mathcal{R} \rangle$ should represent the links in the original *data graph* $\mathcal{G}_d$, and relations $\mathcal{R}$ should represent similarities or relations between links in the original data graph. More formally, let $\mathcal{G}_d \triangleq \langle \mathcal{V}, \mathcal{E} \rangle$ denote a data graph where $\mathcal{V}$ is the set of vertices, and $\mathcal{E}$ is the set of edges or links. In a multi-relational network, vertices and edges can be of different types. For example in a drug-target interaction network, $\mathcal{V}$ is the set of all drugs and protein targets and $\mathcal{E}$ is the set of all the drug-target interactions as well as similarities with different semantics between drugs and between targets. Similarities can be extracted from multiple sources, for example, based on chemical structures of the drugs, or nucleotide sequence of the targets [2].

In such settings, $\mathbf{X}$ is a subset of $\mathcal{E}$, and $\mathcal{R}$ is derived from $\mathcal{G}_d$ based on a modeling decision. For example, Kashima et al. [15] use the *Kronecker sum and product* to derive similarities between links, and Fakhraei et al. [2] use triadic closure principles and define the similarities between links

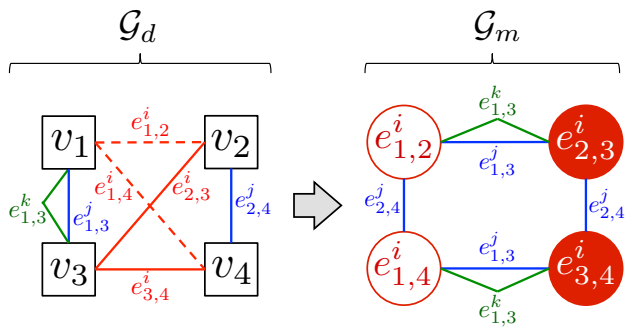

Figure 2: *candidate graph* construction based on the *data graph*. Using the logic shown in (9), the similarity between two links that share a node are based on the similarity between their other nodes. For example, similarity between $\mathbf{x}_p = e_{1,2}^i$ and $\mathbf{x}_q = e_{2,3}^i$ that share $v_2$ is based on the similarities of $v_1$ and $v_3$ which are $e_{1,3}^j$ and $e_{1,2}^k$.

based on the similarities between their end nodes as the following:

$$\text{SIMILAR}(v_2, v_3) \wedge \text{LINK}(v_1, v_2) \Rightarrow \text{LINK}(v_1, v_3) \quad (9)$$
$$\text{SIMILAR}(v_2, v_3) \wedge \neg\text{LINK}(v_1, v_2) \Rightarrow \neg\text{LINK}(v_1, v_3)$$

This model achieves state-of-the-art performance in various domains such as drug-target interaction prediction [2], drug-drug interaction prediction [6], and hybrid recommender systems [7]. Figure 2 shows an example of creating a candidate graph for links based on the *data graph*. In this example, the original data graph has three types of edges ($e^i, e^j, e^k$) and we are interested to infer the values of $e_{1,2}^i$ and $e_{1,4}^i$ based on the other observed edges. In this setting, both nodes and relations in the candidate graph are based on the edges of data graph. e.g., $\mathbf{x}_p = e_{1,2}^i$, $\mathbf{x}_q = e_{1,4}^i$ and $r_{\mathbf{x}_p, \mathbf{x}_q}^j = r_{e_{1,2}^i, e_{1,4}^i}^j = e_{2,4}^j$.

For $\mathcal{G}_n$ in the link prediction setting, we define $\mathcal{I}(x_\alpha, r_\beta)$ as

$$\mathcal{I}\left(\mathbf{x}_\alpha = e_{s,t}^i, r_\beta^j = e_{u,v}^j\right) = \begin{cases} 1 & \text{if } \{s,t\} \cap \{u,v\} \neq \emptyset \\ 0 & \text{otherwise} \end{cases} \quad (10)$$

Link instance $\mathbf{x}_\alpha$ is incident to $r_\beta^j = e_{u,v}^j$ if it has an end point at either $u$ or $v$. Intuitively, links are incident to relations via the nodes connected by the relation.

## 7. EXPERIMENTAL VALIDATION

In this section we present preliminary results of the main components of our framework, activation and nomination methods, on a link prediction dataset. In this dataset we have two types of nodes (drugs and targets), and the task is given the similarities between nodes and partially observed interactions (i.e, links) between them, to predict the held out set of interactions in the network. We use 10-fold cross validation for our experiments where we hold out 10% of the observed links (i.e., positive class) and use the rest as observed instance to predict their values. We also samples 10% of the absent or missing links (i.e., negative class) and include them in each held out fold. Due to high class imbalance in link prediction tasks, the most informative performance measure is the precision and recall of the minority positive class (the presence of link). Therefore, we evaluate our methods based on Area Under the Precision-Recall Curve (AUPR).

The following sections describes the dataset and present our primarily results on it.

### 7.1 Dataset

In this dataset we want to predict new interactions between drug compounds and target proteins. We follow the link predictionmodeling approach described in Section 6 and proposed by Fakhraei et al. [2]. We use known interactions and biologically relevant similarity relations to predict held-out interactions. We describe the interactions and similarities used for our experimental evaluation below.

#### 7.1.1 Drug-Target Interactions

The interactions between drugs and target are gathered from Drugbank, KEGG Drug, Drug Combination Database (DCB) and Matador. The dataset includes 1,306 known interactions between 315 drugs and 250 targets. We use five types of similarity between each pair of drugs and three types of similarity for each pair of targets. We describe each briefly below. For full details, refer to [2]. In this dataset the ratio of positive class (i.e., links presence $y = 1$) to negative class (i.e., link absence $y = 0$) is 1.6%.

Between drug similarities for this dataset include the following: *Chemical-based* relations obtained using the chemical development kit (CDK) and compare the chemical structure of the drug molecules, *Ligand-based* relations computed with the similarity ensemble approach (SEA) search tool and measure the closeness between protein-receptor families for each drug, *Expression-based* relations obtained from the Connectivity Map Project and compare gene expression levels in response to the administration of each drug, *Side-effect-based* relations acquired from the SIDER database and compare the reported side-effects for each drug, and *Annotation-based* relations from the World Health Organization ATC classification system that compares ontological characterizations of drugs.

Between target similarities for this dataset include *Sequence-based* relations computed using the Smith-Waterman sequence-alignment procedure and measure the goodness of alignment between the genetic codes of each target, *Protein-protein interaction network-based* relations computed using the protein-protein interaction network in humans and compare the graph distance between proteins encoded by each target gene, *Gene Ontology-based* relations obtained by downloading Gene Ontology annotations from UniProt and compare the semantic similarity between genes based on their ontological classification.

### 7.2 Results

We use a baseline of selecting $k$ relations or similarities for all instances, and increasing the k at each step. It is important to note that this baseline does not have a nomination quota and basically nominates all instance to receive more relations at each step. Our nomination method in contrast is limited by a quota an can not explore the space as freely as the selected baseline. Figure 3a depicts the performance of only the average weighted Lagrange multiplier (AWL) nomination method with quota of 10% in comparison with the baseline on the drug-target interaction dataset. In this setting, limitation imposed on the search space by AWL nomi-nation method improves the link prediction performance. It is also notable that number of selected similarities at each step by using the nomination method is less than the baseline. Figure 3b shows the performance of only the activation method in comparison to the baseline method. In this setting although all instance were nominated to get more relations, the relations with higher activation score were prioritized to be included in the neighborhood graph, the performance achieved via this method is even higher than the nomination method. It is also notable that the number of similarities selected in this experiment is less than the number of similarities of the baseline method in Figure 3a, which can suggest limiting the search space based on the relation can be more restrictive.

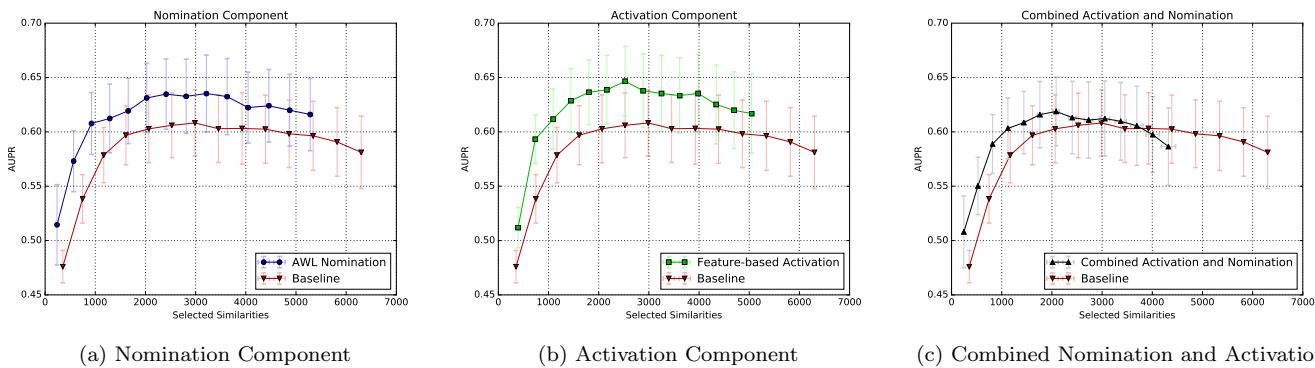

Figure 3: Drug-Target interaction prediction experiments. Horizontal axis show mean and standard deviation (std) of the number of similarities included in the neighborhood graph at each iteration and vertical axis shows the mean and std of the AUPR at each step over ten folds.

The nomination method focuses the search by prioritizing the instance to get more relations, while the activation method directs the search by prioritizing which relations to be selected for the neighborhood graph. Figure 3c shows using combination of nomination and activation methods together, where it achieves higher performances with less numbers of similarities in the beginning and reduces the number of selected similarities in later iterations

## 8. DISCUSSION AND CONCLUSION

In this paper, we highlight the limitations of sequential neighborhood graph construction and introduce a general unified framework to dynamically construct multi-relational neighborhood graphs during inference. We base our dynamic neighborhood graph construction on the states of variables from intermediate inference results, the structural properties of the relations connecting them, and weight parameters learned by the model. We then formulate the general link prediction task as inference on neighborhood graphs, and present initial results on a drug-target interaction network showing effectiveness of our methods. In future work, we plan to extend our studies with experiments on more datasets with various characteristics, experiments with different parameter learning setups, and considering additional methods for the components of our framework such as value-based and probabilistic nomination methods, and label and other structural-based activation methods.

## Acknowledgements
This work is partially supported by the National Science Foundation (NSF) under contract number IIS0746930. Any opinions, findings, and conclusions or recommendations expressed in this material are those of the author(s) and do not necessarily reflect the views of the supporting institutions.